%% file: eprint.tex
\documentclass[10pt]{article}
\usepackage{graphicx}
\def\Title#1{\begin{center} {\Large #1 } \end{center}}
\def\Author#1{\begin{center}{ \sc #1} \end{center}}
\def\Address#1{\begin{center}{ \it #1} \end{center}}

\newcommand\pubblock{\rightline{\begin{tabular}{l} Proceedings of the CTD/WIT 2019\\ \pubnumber\\
         \pubdate  \end{tabular}}}

\newenvironment{Abstract}{\begin{quotation} \begin{center} 
             \large ABSTRACT \end{center}\bigskip 
      \begin{center}\begin{large}}{\end{large}\end{center} \end{quotation}}

\newenvironment{Presented}{\begin{quotation} \begin{center} 
             PRESENTED AT\end{center}\bigskip 
      \begin{center}\begin{large}}{\end{large}\end{center} \end{quotation}}

\input econfmacros.tex
\usepackage{color}
\usepackage{lineno}
\usepackage{subfig}
\usepackage{hyperref}

\newcommand\pubnumber{ PROC-CTD19-094}

\newcommand\pubdate{\today}

\def\affiliation{
Department of Physics and Astronomy \\
Uppsala University, Sweden}

\def\support{\footnote{Work supported by Knut and Allice Wallenberg Foundation}}

\begin{document}

% uncomment the following line for adding line numbers
% \linenumbers

% large size for the first page
\large
\begin{titlepage}
\pubblock

%% Change the title, name, abstract
%% Title 
\vfill
\Title{Time-based Reconstruction of Hyperons at PANDA at FAIR}
\vfill

%  if you need to add the support use this, fill the \support definition above. 
%  \Author{FIRSTNAME LASTNAME \support}
\Author{Jenny Regina $^{a, *}$ \support, Walter Ikegami Andersson $^{a}$}

\center{On behalf of the PANDA Collaboration}
\Address{$^{a}$ \affiliation}
\vfill

\begin{Abstract}
The upcoming PANDA (anti-Proton ANnihilation at DArmstadt) experiment at FAIR (Facility for Anti-proton and Ion Research) offers unique possibilities for performing hyperon physics such as extraction of spin observables. Due to their relatively long-lived nature, the displaced decay vertices of hyperons impose a particular challenge on the track reconstruction and event building. 

The foreseen high luminosity and beam momenta at PANDA requires new advanced tracking algorithms for successfully identifying the hyperon events. The purely software based event selection of PANDA puts high demands on the online reconstruction. A fast, versatile, modular and dynamic approach to track reconstruction and event building is required. This text addresses the reconstruction algorithms used in the scheme such as the Cellular Automaton. A method for obtaining the z-component of the particle momentum is described as well as methods for merging the information from different track reconstruction detectors. 
\end{Abstract}

\vfill

% DO NOT CHANGE!!!
\begin{Presented}
Connecting the Dots and Workshop on Intelligent Trackers (CTD/WIT 2019)\\
Instituto de F\'isica Corpuscular (IFIC), Valencia, Spain\\ 
April 2-5, 2019
\end{Presented}
\vfill
\end{titlepage}
\def\thefootnote{\fnsymbol{footnote}}
\setcounter{footnote}{0}
%

% normal size for the rest
\normalsize 

%% Your paper should be entered below. 

\section{Introduction}
\label{intro}

The PANDA (anti-Proton ANnihilation at DArmstadt)~\cite{PANDAphysicsbook} detector is a novel, almost full 4$\pi$ detector which is currently under construction at FAIR (Facility for Anti-proton and Ion Research). The modular detector will be positioned at the HESR (High Energy Storage Ring) where anti-protons are accelerated and impinge on an internal proton target. PANDA will be one among a new genereation of experiments to utilize a fully software based event selection. For this, a full track and event reconstruction has to be performed in real time at event rates of up to 20 MHz. This puts high demands on the online track reconstruction and event building algorithms. The track reconstruction must be accurate with good momentum resolution, 3-4 $\%$ without a Kalman filter, and vertex determination. It must also be fast in order not to loose data. The exact numbers for this will depend on the final computing resources used for the experiment.

%Hyperons are baryons containing at least one strange quark or one charmed quark in addition to up or down quarks.

\section{The PANDA Experiment}
\label{sec:PANDA}

\begin{figure}[!htb]
  \centering
  \includegraphics[width=0.7\linewidth]{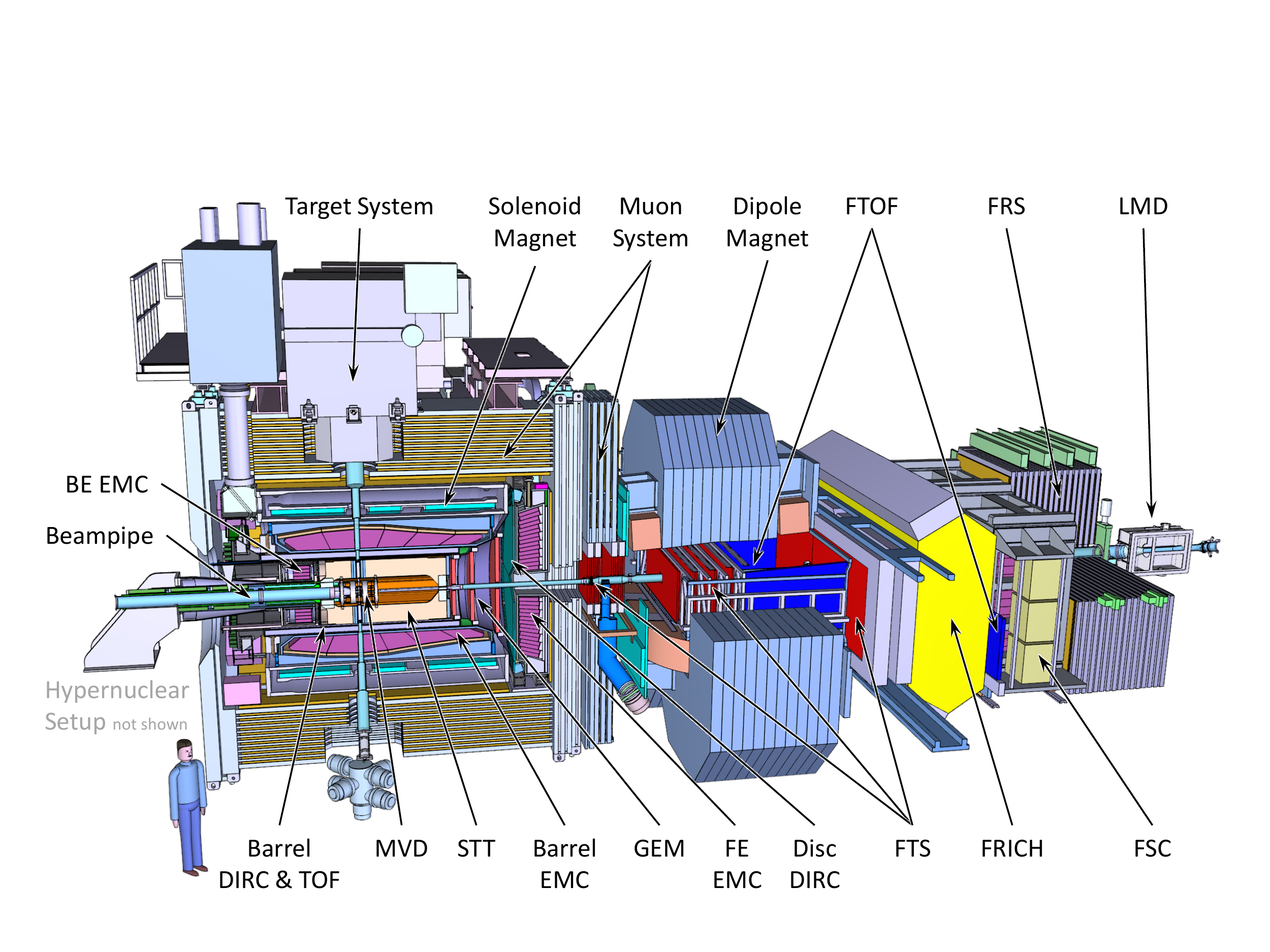}
  \caption{The PANDA detector setup.}
  \label{fig:picture}
\end{figure}

The PANDA detector setup is shown in Figure~\ref{fig:picture}. The anti-proton beam is coming from the left in the figure and impinges on the proton target. The detector setup is divided into two parts, the target spectrometer surrounding the interaction point in a cylindrical shape and the forward spectrometer placed downstream along the beam line. Both parts are needed to provide acceptance for particles emitted at all angles. 

Track reconstruction at PANDA is based on the principle of particle trajectories bending in a magnetic field, a solenoid field in the target spectrometer leading to helix shaped trajectories in 3D and a dipole field in the forward spectrometer leading to hyperbolic trajectories. Track reconstruction and event building is done in the ROOT~\cite{ROOT} based framework of PANDA called PandaRoot~\cite{PandaRoot}.

The main dedicated track reconstruction detector of PANDA is the Straw Tube Tracker (STT)~\cite{STT}. Placed in the target spectrometer, it consists of 4,224 closely packed single channel read out drift tubes with a radius of 0.5 cm and an anode wire in the center. The tubes will contain an $Ar+10\% CO_2$ mixture with a xy-resolution of $\sim$ 0.1 mm. The STT consists of 15-19 layers of straw-tubes arranged parallel to the beam line and the magnetic field lines for track reconstruction in the xy-plane. In addition it has eight layers of tubes skewed by $\pm 3^{\circ}$ for z reconstruction.

A vertex detector, the Micro Vertex Detector (MVD) will be placed closest to the interaction point~\cite{MVD}. It consists of four radial layers, the two innermost consist of silicon pixel detectors and the two outermost consist of silicon strip detectors. In order to cover forward angles as well, six circular planes will be placed downstream in the direction of the beam, four of which will consist of silicon pixel detectors and two of which will consist of a mix of silicon pixel and strip detectors. A hit resolution better than 30 $\mu$m and a vertex resolution of about 100 $\mu m$ is achieved.

\begin{figure}[!htb]
  \centering
  \subfloat[]{\includegraphics[width=0.259\linewidth]{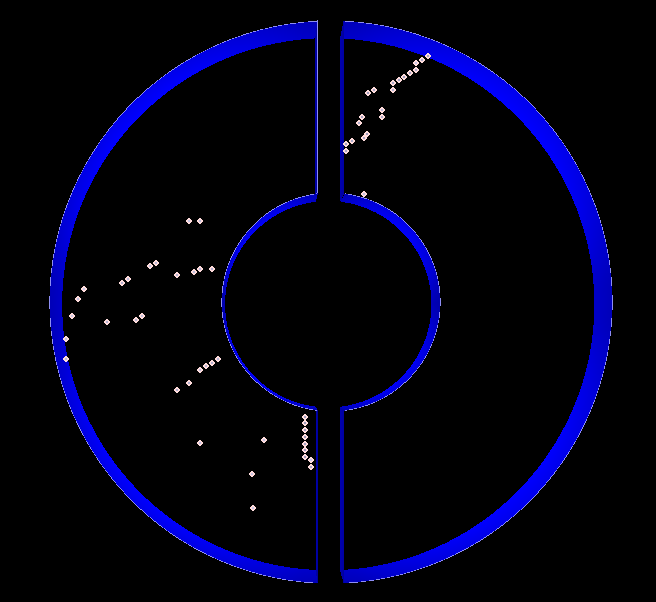}}
  \qquad
  \subfloat[]{\includegraphics[width=0.25\linewidth]{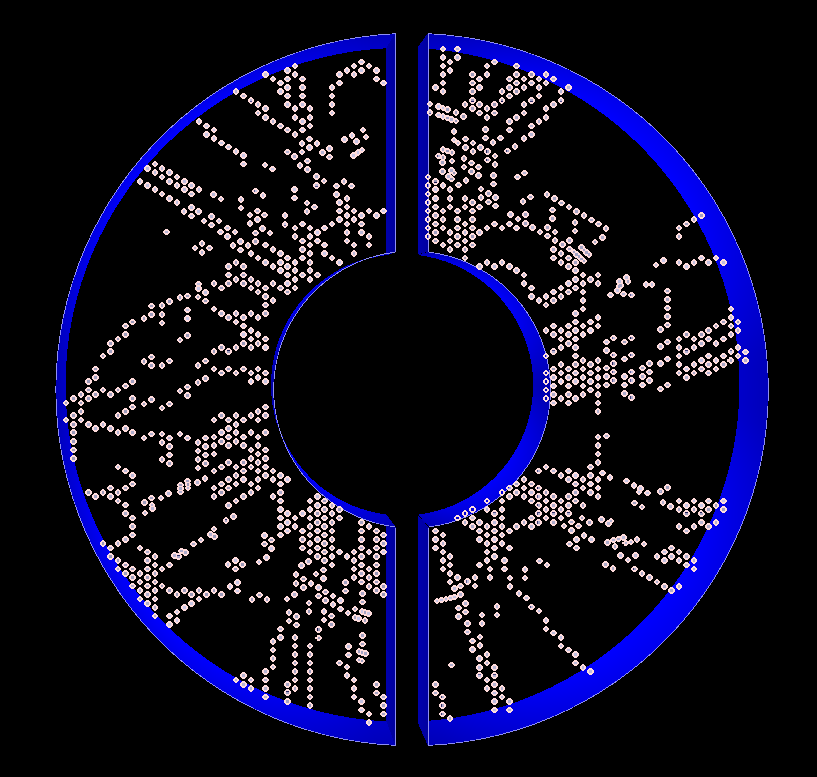}}
  \caption{a) Hits in the STT corresponding to hits less than 200 ns appart in time. b) Hits from overlapping events within 2,000 ns appart in time.}
  \label{fig:pictures}
\end{figure}

At the full luminosity of $\mathcal{L}=2\cdot 10^{32} cm^{-2}s^{-1}$, PANDA will operate at a mean interaction rate of 20 MHz. The HESR will provide a quasi-continuous proton beam with a revolution time of $\sim$ 2,000 ns and a gap of $\sim$ 400 ns. The detector information will be processed in bunches and the beam structure provides a natural time interval of $\sim$ 2,000 ns for these bunches. The case of hits from one event in the STT is shown in Figure~\ref{fig:pictures} a). Bunch of hits occuring within a time-window of 2,000 ns is shown in Figure~\ref{fig:pictures} b). This corresponds to an overlap of four-five events on average. 

% set the height or width as you prefer

\section{Hyperons}
\label{Hyperons}
Hyperons are baryons containing at least one strange quark in addition to the up  and down quark. By investigating the production mechanics of these baryons in $\bar{p}p$ collisions one can probe QCD (Quantum ChromoDynamics)~\cite{QCD} at intermediate energy scales. The long life-time, of the order of $10^{-10}$ s, is a result of a weak decay, which is needed to reduce strangness. This weak decay is self-analysing, hence, it allows us to measure the hyperon's spin. Gathering large data on spin observables enables the test of CP violation, which is expected for the observed matter-antimatter asymmetry. 

Due to their long lived nature the event topology will include one or more displaced vertices which poses a challenge for the reconstruction. They are reconstructed from their decay products, \textit{e.g.}, in the $\bar{p}p\rightarrow\bar{\Lambda}\Lambda\rightarrow p\pi^-\bar{p}\pi^+$ reaction, the $\Lambda$ and $\bar{\Lambda}$ hyperons are identified by the final state particles. Many heavier hyperons decay via lighter neutral hyperons, \textit{e.g.} in the most common decay of the $\Omega^-$, $\Omega^-\rightarrow\Lambda K^-\rightarrow p\pi^- K^-$, where the neutral hyperon leaves no signal in the detector. The specific kinematic distribution of the $\Lambda\bar{\Lambda}$ process, which has previously been thurougly studied at LEAR~\cite{LEAR}, often leaves the $\Lambda$ at rest after scattering. Hence, its decay products tend to have a very low momentum which makes them very difficult to reconstruct.

\section{The SttCellTrackFinder}

One candidate for performing the online track reconstruction at PANDA is called the SttCellTrackFinder~\cite{Bachelor} and utilizes mainly parallel STT tube hit information. The displaced decay vertices of hyperon events require a track finder that does not use the beam-target interaction point as a constraint. A procedure based on the Cellular Automaton~\cite{CellularAutomaton} fulfills this demand. Parts of the track reconstruction algorithm has been paralelized and run on GPUs~\cite{Master}.

The Cellular Automaton based track finding works in the following way. In the first stage all hit tubes, which only have one or two direct neighbors, are selected. These so called unambiguous hits have to come from just one particle track. Each of these tubes get a unique identifier. Each tube compares its id with those of the hit neighbors and takes the lowest one. After a series of iterations, all hits belonging to one track have the same id and form a tracklet. In the second stage all hits with more than two fired neighbors copy the id of their neighbors into an array. At the end the array contains the id of all tracklets which are connected. To select the tracklets which belong together and to obtain the track parameters a circle fit based on a projection on a Riemann surface is done.

Figure~\ref{fig:pictures} shows that at high intensity scenarios or large event bunches unwanted neighbours are created that impede the functionality of the algorithm. A solution to this is to consider the time information of the hits. The current requirement is that two hits need to be no more than 250 ns apart. This is to account for the maximum drift time of the electrons in the tubes and to have some margin to not exclude a correct combination of hits.

Results from first benchmarks of the execution time for different sizes of the event bunces are shown in Table~\ref{tab:timing} which gives the execution time for the part of the code establishing neighborhood relations for 10,000 events. Two different operational modes of the SttCellTrackFinder were used, the time-based where time information is taken into account and the event-based where this information is not utilized. Two different bunch sizes of data for processing were used, bunch sizes of one event and a bunch size where all detector information occuring within 2000 ns is processed. The latter corresponds to a realistic scenario, see section~\ref{sec:PANDA}. The results show a 14 $\%$ relative increase in execution time if the time based option is used and further a 44 $\%$ relative increase in processing time if realistic event bunces are processed compared to the event based case.

%%%%%%%%%%%%%%%%%%%%%%%%%%%%%%%%%%%%%%%%%%%%%%%%%%%%%%%%%%%%%%%%%%%%%%%%%

\begin{table}[!htb]
  \begin{center}    
  \caption{Timing benchmarks for the part of the track reconstruction establishing the hit neighborhood relations.}
    \begin{tabular}{l|ccc}
      \hline
      \hline
      Option for Find Hit Neighbors &  Event Based & Time based & Time based \\
            \hline
      Data bunch size & One event & One event &  2,000 ns \\
      \hline
      Execution Time for 10000 events [s] &  5.09  &  5.80 & 7.31 \\
      \hline
      \hline
    \end{tabular}
    \label{tab:timing}
  \end{center}
\end{table}

%%%%%%%%%%%%%%%%%%%%%%%%%%%%%%%%%%%%%%%%%%%%%%%%%%%%%%%%%%%%%%%%%%%%%%%%%%%

\subsection{Reconstructing the z-component}

\begin{figure}[!htb]
  \centering
  \subfloat[]{\includegraphics[width=0.24\linewidth]{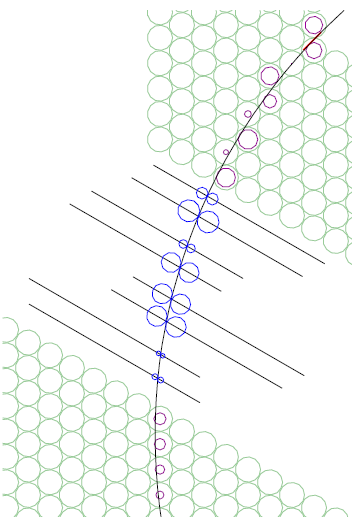}}
  \qquad
  \subfloat[]{\includegraphics[width=0.24\linewidth]{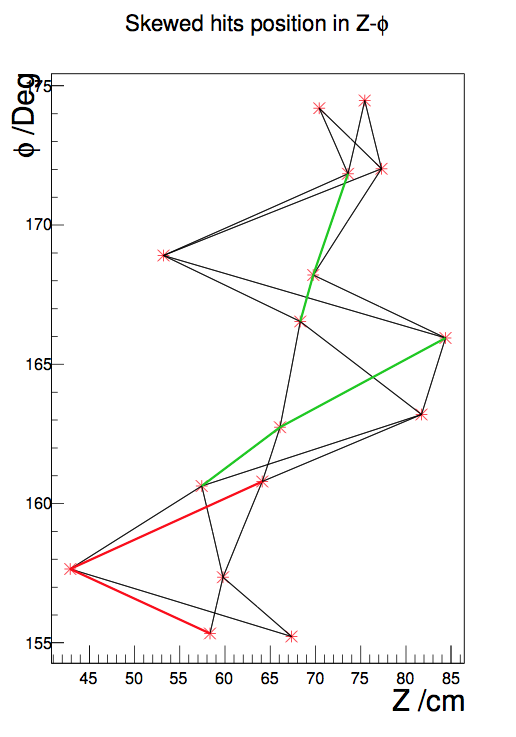}}
    \qquad
  \subfloat[]{\includegraphics[width=0.24\linewidth]{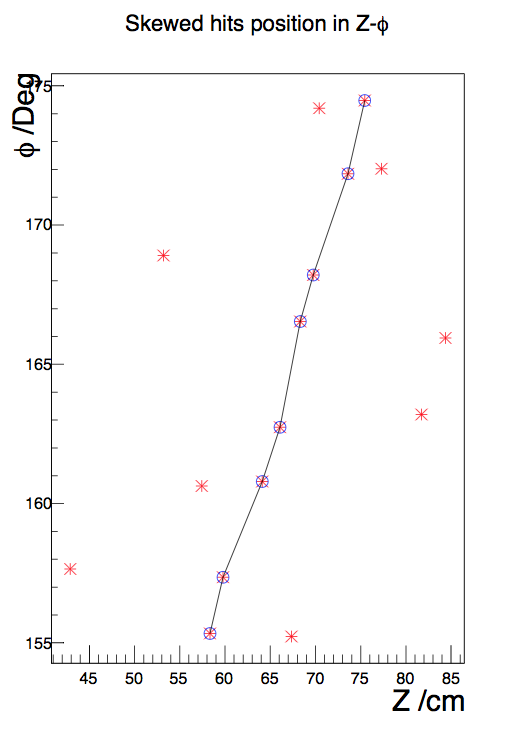}}
  \caption{Panel a) show a segment of the STT and a fitted track in xy-plane. The parallel tubes are displayed as green circles and the drift circles of hit parallel tubes are illustrated as purple circles in the parallel tubes. The xy-projection of skewed straw tubes are shown as black straight lines and the two possible isochrones of skewed straw tubes are illustrated as blue circles. Panel b) show lines drawn between two isochrone centers of adjacent skewed straws. Panel c) show the resulting path.}
  \label{fig:pzFinder}
\end{figure}

The above described procedure works for parallel straw tubes but to include information from the skewed straw tubes, additional algorithms are required~\cite{Walter}. One approach is to use the already fitted track in xy-plane, and search for hits in the skewed tubes whose xy-projection crosses the track.  This is illustrated in Figure~\ref{fig:pzFinder} a). There are two pssibilities for the xy-position of the drift circle center, there is a left/right ambiguity. 

To go on from here, two different approaches have been tested. The drift circle centers can be chosen through a Hough transform procedure or through a combinatorial method. The latter approach is described in the following; the two possible centers of all hit skewed straw tubes are marked in a histogram. Straight lines connecting both possible centers of one drift circle to the centers of drift circles of adjacent tubes are drawn as is illustrated in Figure~\ref{fig:pzFinder} b). This procedure creates paths between the start and the end of the skewed layer. The path with the angle closest to 180$^{\circ}$ between the lines in drift circle centers is chosen to be the most probable one, such a case is illustrated in Figure~\ref{fig:pzFinder} c). In the last step a z-component is obtained from a straight line fit to these center positions.

\subsection{Utilizing Additional Detector Information}

Once a track has been fitted from the STT information, the track reconstruction can be extended to include information from additional detectors. The MVD offers the most precise spatial information for vertex reconstruction. Hence, an algorithm to include its information is under development.

\begin{figure}[!htb]
  \centering 
  \subfloat[]{\includegraphics[width=0.25\linewidth]{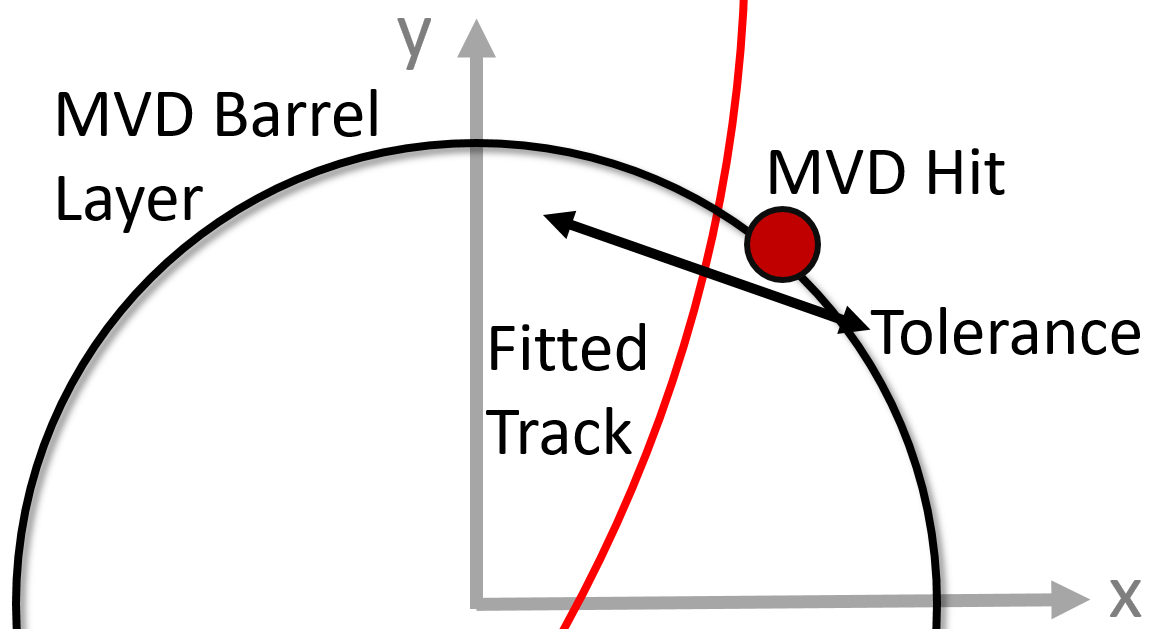}}
    \qquad
  \subfloat[]{\includegraphics[width=0.4\linewidth]{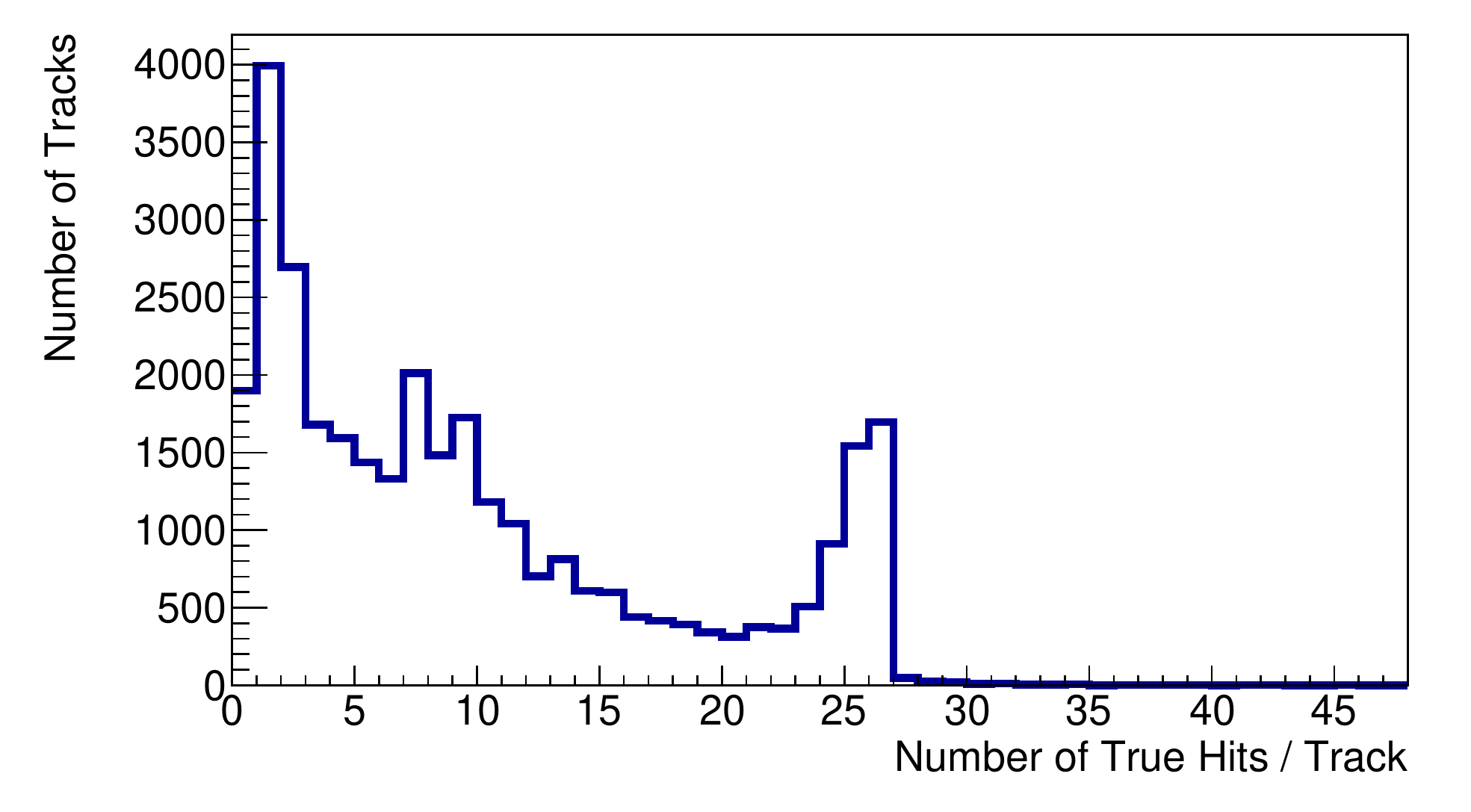}}
  \caption{Panel a) display the method for including MVD hits.}
  \label{fig:MvdSttHits}
\end{figure}

The approach chosen to do this is to use the Riemann track obtained from the STT track reconstruction and the xy-information from the MVD hits. At the moment, hits from the four barrel layers can be included in the track reconstruction algorithm. The search method works in the following way: the outermost layer is first scanned for hits. If one hit is within a certain distance of the track given by the uncertainty of the Riemann track radius, the hit is added to the track and the track is refitted. The refit is performed after every hit inclusion in order for the outermost strip hits to provide a lever arm for the track fitted from the STT and to obtain the new radius error. This is needed for the higher occupancy pixel layers. This method is then repeaded for the rest of the layers from the outside inwards. The algorithm allows for missing hits in layers, if there is no hit within the set distance of the track, no hit from the layer in question is added and the search continues in the next layer. This is illustrated in Figure~\ref{fig:MvdSttHits} a). If several hits are within the set distance of the track, only the closest hit is chosen, therefore, a maximum of four MVD hits can be added to one track. There is no restriction on the number of tracks one single hit can be assigned to at this stage since one wants to allow for intersecting tracks.

One challenging aspect of the inclusion of MVD hits is that it is done in the xy-plane only and the number of hits per event in each pixel barrel layer is quite large so the probability for wrong assignments of hits is quite large as can be seen in Table~\ref{tab:performance}, 16$\%$ of all tracks contain one wrong assigned pixel hit and 37$\%$ contain two wrong assignments. In each silicon strip layer the average number of hits per event is less than two. With this low occupancy the algorithm is more sucessful with 28$\%$ of all tracks containing one wrongly assigned hit but only 17$\%$ containing two. Some work on the fake hit rejection is needed. It should be noted that the total number of tracks considered in this figure are not nessecarily reconstructible. The number of correctly assigned STT hits per track can be seen in figure Figure~\ref{fig:MvdSttHits} b). Only 7 $\%$ of all tracks contain one or more wrongly assigned STT hit and only 1 $\%$ of all tracks contain two or more wrongly assigned hits. This together with Figure~\ref{fig:MvdSttHits} b) leads to the conclusion that the hit purity is rather good.

Tests of the algorithm for xy- STT reconstruction and MVD hit inclusion show a 86$\%$ track finding efficiency for 23,719 simulated tracks. For this number, only simulated tracks with at least six STT hits were considered. A track is considered to be found if the majority of hits belong to the simulated track. Execution time checks for the CPU version of the code on a i7 3.4 GHz processor show that it takes 10 ms/event and STT hit finding takes up approximately 76$\%$ of the runtime and MVD hit finding roughly 24$\%$. A speedup of up to $\times$100 have been achieved on a GeForce GTX 750 Ti GPU for the STT hit finding~\cite{Master}. 

% set the height or width as you prefer

\begin{table}[!htb]
  \begin{center}   
   \caption{Fraction of all tracks containing one or two true or false MVD strip or Pixel hits.}
    \begin{tabular}{l|cccc}
      \hline
      \hline
      Type of Hits  & True MVD Pixel & True MVD Strip & False MVD Pixel & False MVD Strip \\
            \hline
      Number of Hits &  1 / 2  & 1 / 2 & 1 / 2 & 1 / 2 \\
      \hline
      $\%$ of tracks & 2 / 0.7 &  22 / 9 & 16 / 37 &  28 / 17 \\
      \hline
      \hline
    \end{tabular}
    \label{tab:performance}
  \end{center}
\end{table}

%%%%%%%%%%%%%%%%%%%%%%%%%%%%%%%%%%%%%%%%%%%%%%%%%%%%%%%%%%%%%%%%%%%%%%%%%%%
\vspace{-2mm}
\section{Conclusions}

Reconstruction of low momentum tracks and displaced vertices in real time at PANDA is a challenging task. An algorithm which is able to do this is currently under development. The SttCellTrackFinder based on a Cellular Automaton and a Riemann fit show promising results and is a candidate for online track reconstruction at PANDA due to its parallelizability. Tests show a 86 $\%$ track finding efficiency. The algorithm is able to process data sorted according to detector time-stamps and further it runs on GPUs. 

Further tests and optimization will be performed. The scaling with vertex displacement and $p_T$ are to be evaluated. A cleanup and track merging procedure will be developed as well as a fake hit rejection algorithm. Tests on hyperon events with event mixing will be performed.

%%%%%%%%%%%%%%%%%%%%%%%%%%%%%%%%%%%%%%%%%%%%%%%%%%%%%%%%%%%%%%%%%%%%%%%%%%%

\end{document}

%% file: econfmacros.tex
\def\beq{\begin{equation}}
\def\eeq#1{\label{#1}\end{equation}}
\def\eeqn{\end{equation}}

\def\beqa{\begin{eqnarray}}
\def\eeqa#1{\label{#1}\end{eqnarray}}
\def\eeqan{\end{eqnarray}}

\let\bar=\overbar

\def\Dslash{\not{\hbox{\kern-4pt $D$}}}
\def\dslash{\not{\hbox{\kern-2pt $\del$}}}

\def\msb{{\bar{\ssstyle M \kern -1pt S}}}